\documentclass[twocolumn,showpacs,preprintnumbers]{revtex4}
\usepackage{amssymb}
\usepackage[dvips]{graphicx}

\begin{document}

\title{'Giant' normal state magnetoresistances
of Bi$_{2}$Sr$_{2}$CaCu$_{2}$O$_{8+\delta}$.}

\author{ V.N. Zavaritsky$^{1,2}$, J. Vanacken$^{3}$, V.V. Moshchalkov$^{3}$, A.S. Alexandrov$^{1}$}
\address
{$^{1}$Department of Physics, Loughborough University, Loughborough, United Kingdom, 
$^{2}$Kapitza Institute for Physical Problems,  Moscow, Russia,
$^{3}$Laboratory for Solid State Physics and Magnetism K.U.Leuven, Leuven, Belgium}

\begin{abstract}
Magnetoresistance (MR) of Bi-2212 single crystals with T$_{c}$ $\approx 87-92 K$ is studied  in pulsed magnetic fields up to 50T along the c-axis in a wide temperature range. The negative out-of-plane and the positive in-plane MRs are measured in the normal state. Both MRs have similar magnitudes, exceeding any orbital contribution by two orders in magnitude. These are explained as a result of the magnetic pair-breaking of  preformed pairs. Resistive upper critical fields H$_{c2}$(T) determined from the in-- and out-of-plane MRs are about the same. They show non-BCS temperature dependences compatible with the Bose-Einstein condensation field of preformed charged bosons.
\pacs{74.72.Hs, 74.25.Fy, 74.25.Op, 74.20.Mn}
\end{abstract}

\maketitle
In the cuprates \cite{gen,mac,oso,alezav,and,car,fra,zav,gan,zav2}, high magnetic field studies revealed a non-BCS upward curvature of resistive $H_{c2}(T)$. When measurements were performed on low-T$_c$ cuprates and other unconventional superconductors \cite{mac,oso,fra,spin,org}, the Pauli limit was exceeded by several times. A non-linear temperature dependence in the vicinity of T$_{c}$ was unambigously observed in a few samples \cite{alezav,fra,gan,zav2}. This strong departure from the canonical BCS behaviour led some authors \cite{car,wen,luo} to conclude, that the abrupt resistive transition in applied fields is not a normal-superconductor transition at $H_{c2}$, and superconductivity could survive in the $CuO_2$ layers well above the resistive $H_{c2}(T)$ line.

The apparent controversy in the different determinations of H$_{c2}(T)$ needs to be further addressed both experimentally and theoretically. In particular the bipolaron theory \cite{ale,alemot} suggests that unconventional superconductors could be in the 'bosonic' limit of preformed real-space pairs, so their resistive $H_{c2}$ is actually a critical field of the Bose-Einstein condensation of charged bosons \cite{ale,zavkabale}. On the experimental side it was particulary important to verify that $H_{c2}$ determined from c-axis and in-plane resistivities data yield the same value especially for extremely anisotropic medium.
High magnetic field studies also revealed the
negative c-axis longitudinal magnetoresistance (MR)
\cite{c-axis,zav,zav2} above $T_{c}$ while the
in-plane transverse MR was found to be positive
\cite{in-plane}. Because of its 'normal' sign, several authors
attributed the in-plane MR above $T_c$ to orbital
effects. However these MRs were measured on different samples, so that their quantitative comparison  was not possible in most cases, and their microscopic origin has remained unknown. 

We report here on a study of both, the in-plane and the out-of-plane MRs, $R_{ab}(B)$ and $R_c(B)$, of the same
Bi$_{2}$Sr$_{2}$CaCu$_{2}$O$_{8+\delta}$ (Bi-2212) single crystals subjected to a
pulsed magnetic field up to 50 Tesla along the c-axis. In contrast with the negative out-of-plane MR, the normal state in-plane MR is found to be positive. Despite the opposite sign of the MRs and huge anisotropy ($\rho_c/\rho_{ab}\gg10^4$), relative magnitudes of the MRs, $\delta R(B)/R(0)$ {\it and} their temperature dependences are found to be similar. Qualitatively and quantitatively similar estimates of the upper critical field $H_{c2}(T)$ were obtained from the in-- and out-of-plane data taken at $T<T_c$. $H_{c2}(T)$ shows a divergent behaviour consistent with  results obtained in other materials \cite{mac,oso,alezav,fra,boe,zav,gan}.  We propose a microscopic explanation of these observations  based on the bipolaron theory of cuprates.

 Bi-2212 single crystals were grown by solid state reaction
\cite{zav} and had a zero-field transition temperature,
T$_{c}\approx87-92K$. 
We measured $R_c$ on samples with in-plane dimensions from $\simeq 80 \times 80 \mu m^2$ to $\simeq 30 \times 30 \mu m^2$ while $R_{ab}$ was studied on a longer crystals, from $\simeq300\times11\mu m^2$ to $\simeq780\times22\mu m^2$.  All samples for the in-plane and out-of-plane measurements were cut from the same parent crystals of $1-3\mu m$ thickness. The zero-field in-- and out-of-plane resistances, $R_c(T)$ and $R_{ab}(T)$, typical for the six pairs of samples selected for this study, are shown by solid lines in the inset to Fig.1. Metallic type of $R_{ab}(T)$ indicates vanishing out-of-plane contribution to $R_{ab}$. Each crystal was fixed on a quartz substrate with four leads made of $5\mu m$ gold wire; the misorientation between the field and the c-axis of the crystal was estimated to be less than a few degrees. No detectable change in the orientation was observed even after a sequence of numerous 50T-pulses of different polarity. However, noticeable {\it reduction} of a zero field $R_{ab}(T)$ was observed on two samples during extensive investigations with 50T-pulses. This change became measurable after 30-60 pulses and reached quasi-stable value after similar number of shots. The `final' $R_{ab}(T)$ is shown in the inset to Fig.1 by the long dashed line. Field induced modification of the inhomogeneities' distribution might be responsible for the effect which requires further investigation.  

The absence of hysteresis in the data obtained on the rising and falling sides of the pulse and the consistency of measurements made at the same temperature in pulses of different $H_{max}$ excludes any measurable eddy-current heating effects.  This is confirmed by a consistency of dc-R(B) taken at identical conditions on  different currents, $j=10-1000A/cm^2$ for $R_{ab}$ and $0.1-20A/cm^2$ for $R_c$. 
\begin{figure}
\begin{center}
\includegraphics[angle=-0,width=0.47\textwidth]{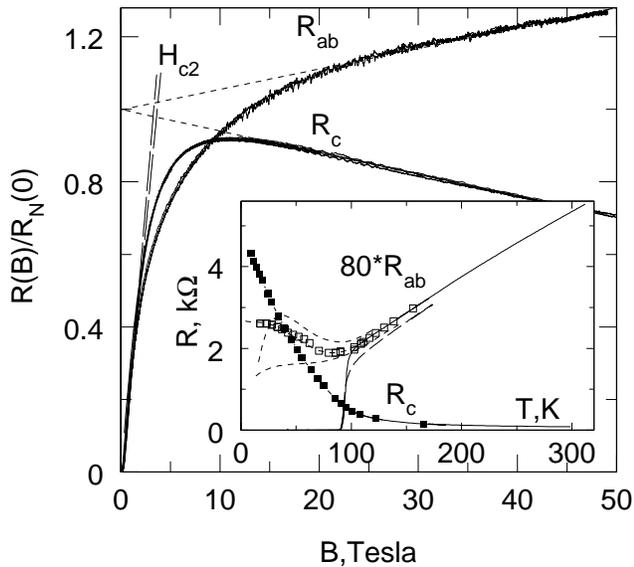}
\caption{$R_c(B)$ and $R_{ab}(B)$ of Bi-2212 at  $\simeq70K$ normalised by $R_N(0,T)$ obtained with the linear extrapolation from the normal state region (short dashes). The linear fits, shown by long dashed lines, refer to flux-flow region. The inset shows the zero-field $R(T)$ (solid lines) together with estimates of $R_N(0,T)$ marked by symbols explained in the text.}
\vspace{-2.5mm}
\end{center}
\end{figure}

Fig.1 shows a typical effect of magnetic field on $R_c(B)$ and $R_{ab}(B)$ resistances of a Bi-2212 single crystal below $T_{c}$. The low-field portions of the curves are attributed to the resistance driven by vortex dynamics.  The thermally activated flux flow is responsible for a nonlinear power-law field dependence, which is followed by a  regime, $R_{FF}(B,T)$, where a linear field dependence fits the experimental observations rather well, Fig.1. We ascribe this positive linear MR, which persists to the lowest temperatures to the flux-flow, $R_{FF}(B,T)$. Flux-flow resistance would not normally be expected in a longitudinal geometry, $R_c(B)$, but a highly anisotropic structure, with alternating quasi-metallic and disordered non-metallic layers, would favour current paths with in-plane meanders, leading to a finite Lorentz force applied to the vortex \cite{zav2}.  It is natural to attribute the high field portions of the curves in Fig.1 (assumed to be above H$_{c2}(T)$) to a normal state. Then, in agreement with original findings of \cite{zav,zav2}, the c-axis high-field MR 
appears to be negative and quasi-linear in B in a wide temperature range both above and below $T_{c}$. Contrary to $R_c(B)$, normal state in-plane MR is positive as seen in the insert to Fig.2 (we use unified notations  in the figures throughout the paper, where the open and  solid symbols refer to $R_{ab}$ and $R_{c}$ respectively). Rapid change of its relative value with temperature may be fitted by $|\delta R(B)/R(0)|\propto (T_1/T)sinh(T_s/T)$ as shown by the solid line in the main panel of Fig.2 with $T_1=0.6K$ and $T_s=320K$. As it is clearly seen from Fig.2, both, the in-plane and the out-of-plane MR in the normal state have similar magnitudes and temperature dependences. The giant magnitudes of MRs, which are about or larger than 1\%, Fig.2, rule out their orbital origin. This is because the measured Hall angle in Bi-2212, is rather small, $\Theta_H <10^{-3}B$ ($B$ in Tesla), \cite{hallangle}, so that the orbital contribution to MRs is less than $0.2$\%. Dashed line in Fig.2 shows re-scaled low-field data reported previously in \cite{in-plane} for a whisker of significantly lower $T_c\approx69K$. Evident similarity between dependences in this panel suggests a normal state origin of the effect and its universality \cite{leuven}. 
\begin{figure}
\begin{center}
\includegraphics[angle=-0,width=0.47\textwidth]{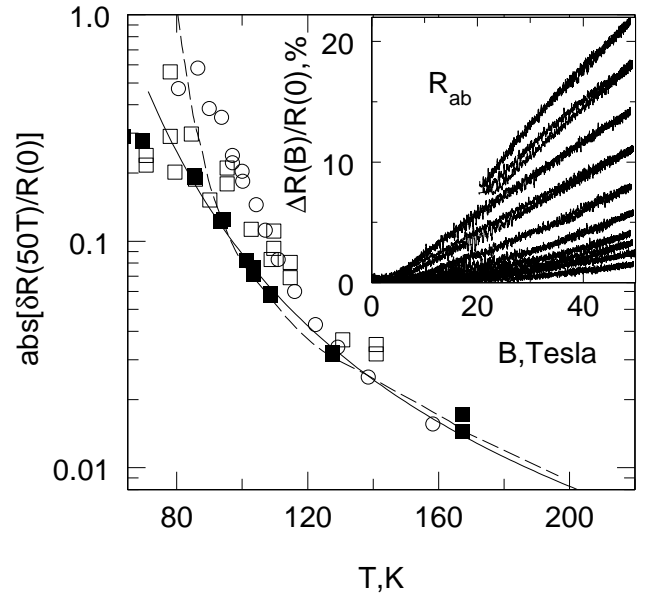}
\vskip -0.5mm
\caption{The {\it absolute} value of in-- and out-of-plane normal state MR taken at 50T, $[R(50T)-R(0)]/R(0)$. Solid line shows the theoretical fit; the dashed line is drawn through rescaled low-field data of the whisker of lower $T_c$\cite{in-plane}. The inset shows in-plane $\delta R(B)/R(0)$ taken  at T$\sim$ 97, 100, 104.2, 107, 111, 116, 122.4, 129.2, 138.5, and 158.2K (from the top).}
\vspace{-2.5mm}
\end{center}
\end{figure} 

Zero-field resistance of the crystals in the absence of superconductivity, $R_N(0,T)$, estimated for in-- and out-of-plane transport by the extrapolation of the high-field portion of experimental $R(B)$ to $B=0$ are shown in the insert to Fig.1. There is some uncertainty in the estimates of $R_N(0,T)$ growing up on temperature lowering due to shrinking of the range of total suppression of the superconductivity by experimentally accessible fields. There is also somewhat different in-plane $R_N(0,T)$ for different crystals studied. This non-universality of the $R_N(0,T)$ is illustrated by short dashed lines in the insert. However it should be emphasised that the upper critical field, $H_{c2}(T)$ (see below), is virtually insensitive to the particular choice of $R_N(0,T)$.  

 Referring to Fig.3, the inset shows the field dependence of Bi-2212 in-plane resistance normalised by its normal state zero-field value, $R_{N}$. Very similar result is obtained for $R_c(B)$ as in \cite{zav,zav2}.  A striking difference in comparison with conventional superconductors  is a progressive broadening of the transition with increasing field instead of a nearly parallel shift.  We consider this to be a result of an unconventional shape of H$_{c2}$(T) since the slope of the flux-flow resistance is inversely proportional to $H_{c2}$ as R$_{FF}= R_{N}\times B/H_{c2}$. The unusual  temperature dependence of this slope, $\partial R_{FF}/\partial B$, is clearly seen from the main panel of Fig.3.   
\begin{figure}
\begin{center}
\includegraphics[angle=-0,width=0.47\textwidth]{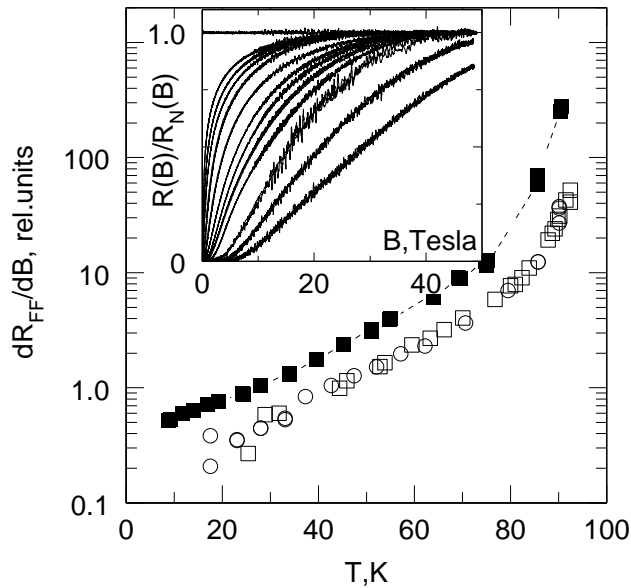}
\vskip -0.5mm
\caption{Flux-flow resistance of Bi-2212, determined from as-measured $R_{ab}$ and $R_c$. The inset shows the in-plane MR normalised by $R_N(B)$ for T$\sim$ 23.1, 28, 33.1, 37.3, 42.7, 47.5, 52.1, 62.1, 70.6, 79.4, 85.7, 90.1, and 108.8K (right to left).}
\end{center}
\vspace{-2.5mm}
\end{figure} 

The resistive upper critical field, H$_{c2}$(T), is estimated from $R_c(B)$ and $R_{ab}(B)$ either as the intersection of two linear approximations in Fig.1, or from the definition of the flux-flow resistance, $H_{c2}=R_N(0,T)(\partial R_{FF}/\partial B)^{-1}$; both estimates are found to be almost identical. This procedure allows us to separate contributions \mbox{originating} from the normal and superconducting states and, in particular, to avoid an ambiguity
due to fluctuations in the crossover region. $H_{c2}(T/T_c)$ estimated from $R_{ab}(B)$ and $R_c(B)$, is shown in Fig.4 together with our estimate of $H_{c2}$ obtained by similar method  \cite{comment} from independent studies of in-- and out-of-plane MR in parent compound,
Bi-2201 of similar anisotropy but of lower $T_c$$\approx$13-25$K$ \cite{zha,and}. Reasonable agreement between $H_{c2}(T)$ estimates from $R_c$ and $R_{ab}$ which is evident from Fig.4 favours our assignment of resistive $H_{c2}$ to the upper critical field especially taking into account the extreme electric anisotropy of the crystals (insert to Fig.4) and presumably different mechanisms responsible for $R_{ab}$ and $R_c$.
\begin{figure}
\begin{center}
\includegraphics[angle=-0,width=0.47\textwidth]{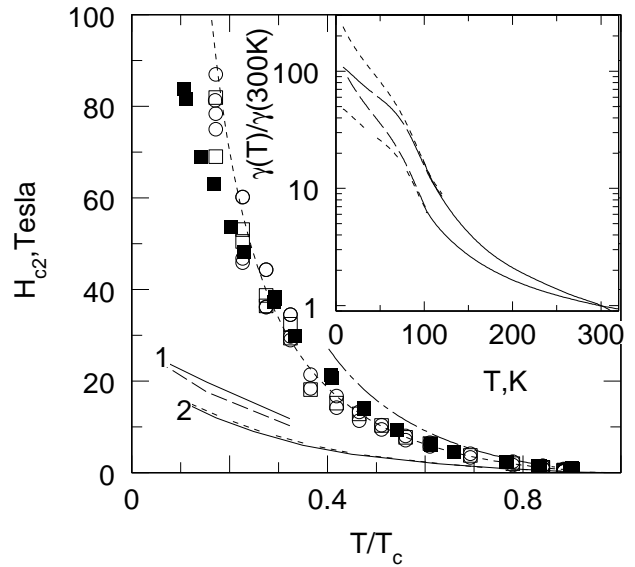}
\vskip -0.5mm
\caption{$H_{c2}$ estimated from the in-- and out-of-plane MR in Bi-2212 are shown together with the fit to Eq(4) (dashed line). Estimates obtained in \cite{comment} from independent studies of Bi-2201 by \cite{and} and \cite{zha} are labelled as 1 and 2 respectively. Broken lines in these pairs  correspond to the data taken from $R_c$, solid - to $R_{ab}$. The inset addresses the anisotropy of our crystals; $\gamma(T)\propto\rho_c(T)/\rho_{ab}(T)$; $\gamma(300K)\approx(2-5)10^4$.}
\end{center}
\end{figure} 

In what follows, we show that the unusual features of the c-axis and in-plane
magnetotransport, Figs.  1-4, can be broadly understood within a phenomenological model of preformed pairs (bosons) in particular, with  bipolarons \cite{ale,alemot}. Within the theory  holes are paired at any temperature into bipolarons, which coexists with thermally excited unpaired carriers. The  c-axis $normal$ state  transport in cuprates is dominated by single unpaired holes, except at low temperatures where unpaired carriers are frozen out \cite{AKM}. This is due to a  large c-axis effective mass of bipolarons compared with the polaron mass. But the in-plane transport is due to both mobile bipolarons and polarons because their in-plane masses are comparable. Single polarons exist as excitations with the energy k$_{B}$T$^{*}$=$\Delta/2$ (the normal state pseudogap) or larger, where $\Delta$ is the bipolaron binding energy.  The edges of two polaronic spin-split bands depend on the magnetic field due to  spin and orbital magnetic
shifts as \cite{zav2}
\begin{equation}
{\Delta_{\downarrow,\uparrow}\over{2}}={\Delta\over{2}}+\mu_{B}^{*}B
\pm (J\sigma +\mu_{B}B),
\end{equation}
where $J$ is the exchange interaction of holes with localised copper electrons and $\sigma$ is an average magnetisation of copper per site ($T_s$=$J\sigma/k_B$ is about hundred K \cite{zav2}). The exchange interaction leads to the spin-polarised polaron bands split by $2J\sigma$.  They are further split (the last term in Eq.(1)) and shifted by the
external magnetic field.  Here $\mu_{B}$ and $\mu_{B}^{*}$ are the Bohr magnetons determined with the electron $m_{e}$ and polaron $m^{*}$ mass, respectively. Assuming that k$_{B}$T is less than the polaron bandwidth and noting that polarons are not degenerate at any temperature,  we obtain for the polaron density
\begin{equation}
n_{p}\sim T^\frac{d}{2}\exp\left(-\frac{T^*}{T}-\frac{\mu_{B}^*B}{k_BT}\right)\cosh\left(\frac{J\sigma+\mu_BB}{k_BT}\right),
\end{equation}
where d is the dimensionality of their energy spectrum.

As follows from Eq.(2) the density of polarons increases with the magnetic field, while the density of bipolarons decreases, if the total number of carriers does not depend on the magnetic field. That explains both the negative c-axis and positive in-plane MR, Fig.1-2. It is reasonable to assume that the (bi)polaron mobility, $\mu_{b,p}=e^*\tau_{b,p}/m_{b,p}$, is field  independent in the relevant region of $B$ because the Hall angle is small. Here $\tau_{b,p}$, $m_{b,p}$ are the relaxation times and  effective masses, respectively, $e^*$=$2e$ for bipolarons and $e^*$=$e$ for polarons. Then the in-- and out-of-plane normal state MRs are given by \begin{equation}
{\delta R_{c,ab}(B)\over{R_{c,ab}(0)}}= -{\delta
n_{p}\over{n_{p}}}{{r_{c,ab}-1}\over{ r_{c,ab}+n_{b}/ n_{p}}}.
\end{equation}
Here $n_b$ is the bipolaron density, $r_{c,ab}=\mu_p^{c,ab}/(2\mu_b^{c,ab})$ is half of the ratio of the polaron and bipolaron mobilities, and $\delta n_{p}$ is a change of the polaron density with the magnetic field. According to the theory \cite{alemot} $r_c>1$, so that the $c-axis$ MR, Eq.(3), is negative. However   the in-plane MR is positive, Eq.(3), if twice the in-plane bipolaron mobility is larger than the in-plane polaron mobility, $r_{ab}<1$. Both relative MRs are of the same order of magnitude, though of the opposite sign. We estimate $\delta n_{p}/n_{p}\simeq (\mu_BB/(k_BT))\sinh(T_s/T)$, which is about 0.1  at $T=100$K and $B=50$T, as observed, Fig.2.

Finally, resistive  $H_{c2}(T)$ determined from the c-axis and in-plane data are virtually the same as seen from Fig.4 where the temperature dependence of $H_{c2}$ is presented together with the  theoretical Bose-Einstein condensation  field \cite{ale} given by
 \begin{equation}
 H_{c2}(T) \sim (t^{-1}-t^{1/2})^{3/2}
 \end{equation}
 with $t=T/T_{c}$. Both $H_{c2}(T)$ show an upward temperature dependence in agreement with Eq.(4).

Our  model of the c-axis and in-plane magnetotransport is supported by other independent observations. In particular, the temperature dependences of the in-plane \cite{BRAT,in,hall} and out-of-plane resistivities \cite{AKM,out}, magnetic susceptibility \cite{AKM,MULLE,hall}, and of the Hall effect \cite{BRAT,hall} strongly support the bipolaron origin of the normal state pseudogap T$^{*}$. The isotope effect on the carrier mass \cite{guo} provides another piece of evidence for (bi)polaronic carriers in cuprates.

 In conclusion, we have  measured the longitudinal out-of-plane and transverse in-plane MR of Bi-2212 single crystals in magnetic fields up to 50 T. We observed a negative c-axis MR and a  positive in-plane MR in the normal state of the same samples  of Bi-2212 and discovered the quantitative similarity of their magnitudes. We determined the resistive upper critical field, which is virtually the same from both resistivities. The opposite sign of out-of-plane and in-plane magnetoresitances, their magnitudes, and the unusual shape of H$_{c2}$(T) were interpreted within the framework of the bipolaron theory.

This work has been  supported by the Leverhulme Trust (UK;F/00261/H) and FWO \& IUAP Programms. We appreciate stimulating discussions with  A.F.~Andreev,  N.E.~Hussey,  V.V.~Kabanov, W.Y.~Liang,   and K.A.~Muller. We are especially grateful to J.R.~Cooper for valuable remarks and for the opportunity to perform zero-field measurements on his dedicated equipment. The authors are thankful to many colleagues at the K.U.Leuven, in particular, to  S.~Stroobants and  T.~Wambecq for their generous help with experiments.


\begin{thebibliography}{99}
\bibitem{gen}  B.Bucher {\it et al}.,
Physica C {\bf \ 167}, 324 (1990).

\bibitem{mac}  A.P.Mackenzie $et$ $al$, Phys. Rev. Lett. ${\bf 71}$, 1238
(1993).

\bibitem{oso}  M.S.Osofsky $et$ $al$, Phys. Rev. Lett. {\bf 71}, 2315
(1993).

\bibitem{alezav}  A.S.Alexandrov $et$ $al$, Phys. Rev. Lett. ${\bf 76}$,
983 (1996).

\bibitem{and}  Y.Ando $et$ $al$, Phys. Rev. Lett. ${\bf 77}$, 2065 (1996).


\bibitem{car}  A.Carrington {\it et al}, 
Phys. Rev. B {\bf 54}, R3788 (1996).

\bibitem{fra}  D.D.Lawrie $et$ $al$, J. Low Temp. Phys. ${\bf 107}$, 491
(1997).

\bibitem{zav}  V.N.Zavaritsky, M.Springford, JETP Lett. {\bf 68},
448 (1998);  V.N.Zavaritsky, JETP Lett. {\bf 71}, 80 (2000).

\bibitem{gan}  V.F.Gantmakher $et$ $al$, JETP {\bf 88}, 148 (1999).

\bibitem{zav2}  V.N.Zavaritsky, M.Springford, A.S.Alexandrov,
Europhys. Lett. {\bf 51}, 334 (2000).

\bibitem{spin}  T.Nakanishi $et$ $al$, Int. J. Mod. Phys. {\bf 14}, 3617
(2000).

\bibitem{org}  I.J.Lee {\it et al}, 
Phys. Rev. B {\bf 62}, R14 669 (2000).

\bibitem{wen}  \mbox{H.H.Wen, S.L.Li, Z.X.Zhao, Phys.Rev. B{\bf 62},
716 (2000)}


\bibitem{luo}  J.L.Luo $et$ $al$, cond-mat/0112065.

\bibitem{ale}  A.S.Alexandrov, Phys. Rev. B${\bf 48}$, 10571 (1993).

\bibitem{alemot}  \mbox{A.S.Alexandrov,N.F.Mott,Rep.Prog.Phys.${\bf 57}$,1197(1994)}

\bibitem{zavkabale} V.N.Zavaritsky, V.V.Kabanov, and A.S.Alexandrov, Europhys. Lett. ${\bf 60}$, 127 (2002).

\bibitem{c-axis}Y.F.Yan {\it et al.}, Phys. Rev. B {\bf52}, R751, (1995).

\bibitem{in-plane}J.R.Harris {\it et al.}, Phys.Rev.Lett. {\bf75}, 1391,
(1995); \\Yu.I.Latyshev{\it et al.}, 
Europhys. Lett. {\bf 29}, 495 (1995).


\bibitem{boe}G.S.Boebinger $et$ $al$, Phys. Rev. B {\bf 60}, 12475 (1999).

\bibitem{hallangle} Z.Konstantinovic $et$ $al.$, Physica B ${\bf 259-261}$, 569 (1999); Phys. Rev. B ${\bf 62}$, R11989 (2000).

\bibitem{leuven} A large MR was also noticed in other cuprates, e.g. in LaSrCuO by J.Vanacken et al., cond-mat/0308227.

\bibitem{comment} V.N.Zavaritsky, M.Springford, A.S.Alexandrov, cond-mat/0011192.

\bibitem{zha} Y.Z.Zhang $et$ $al$, Phys. Rev. B{\bf 61}, 8675 (2000).



\bibitem{AKM}  A.S.Alexandrov, V.V.Kabanov and N.F.Mott, Phys. Rev. Lett. {\bf 77, }4796 (1996).


\bibitem{BRAT}  A.S.Alexandrov, A.M.Bratkovsky and N.F.Mott, Phys. Rev. Lett, {\bf 72}, 1734 (1994)

\bibitem{in}  X.H.Chen {\it et al.}, Phys. Rev. B{\bf 58,} 14219 (1998); W.M.Chen {\it et al.}, 
Physica~C {\bf 341}, 1875 (2000).

\bibitem{hall} A.S.Alexandrov, V.N.Zavaritsky,  S.Dzumanov, cond-mat/0304152.

\bibitem{out}  J.Hofer et al., Physica C {\bf 297}, 103 (1998); V.N.Zverev
and D.V.Shovkun, JETP Lett. {\bf 72}, 73 (2000).

\bibitem{MULLE}  K.A.Muller et al., J. Phys. Cond. Mat., {\bf 10}, L291
(1998).

\bibitem{guo}  G.Zhao et al., Nature, {\bf 385}, 236 (1997).

\end{thebibliography}
\end{document}